# Artificial Intelligence Governance and Ethics: Global Perspectives

Report

28 June 2019

## Authors:

Angela Daly, Thilo Hagendorff, Li Hui, Monique Mann, Vidushi Marda, Ben Wagner, Wei Wang and Saskia Witteborn

This report is supported by Angela Daly's Chinese University of Hong Kong 2018-2019 Direct Grant for Research 2018-2019 'Governing the Future: How are Major Jurisdictions Tackling the Issue of Artificial Intelligence, Law and Ethics?'

The authors acknowledge the research assistance of Ms Jing Bei and Mr Sunny Ka Long Chan, and the comments and observations from participants in the CUHK Law Global Governance of AI and Ethics workshop, 20-21 June 2019.

# Table of Contents

# Author biographies

**Angela Daly, PhD,** is a socio-legal scholar working on transnational regulation of new (digital) technologies. She is the author of *Socio-Legal Aspects of the 3D Printing Revolution* (Palgrave 2016) and *Private Power, Online Information Flows and EU Law: Mind the Gap* (Hart 2016) and co-editor of *Good Data* (Institute of Network Cultures 2019). She holds a PhD in Law from the European University Institute, is currently based at the Chinese University of Hong Kong Faculty of Law, and will join Strathclyde University Law School (Scotland) as Senior Lecturer (Associate Professor) in September 2019. She holds visiting and adjunct positions at Queensland University of Technology (Australia), Tilburg University (Netherlands) and University of Macerata (Italy).

**Thilo Hagendorff, PhD,** is a media and technology ethicist. He received his doctorate in 2013 with a sociological thesis. He has been a research associate at the International Center for Ethics in the Sciences and Humanities (IZEW) at the University of Tuebingen (Germany) since 2013. Since 2019 he is working in the 'Machine Learning: New Perspectives for Science' Excellence Cluster. Furthermore, he is a lecturer at the University of Tuebingen as well as at the Hasso-Plattner-Institute of the University of Potsdam (Germany).

**Li Hui, PhD,** is an associate research fellow at the Shanghai Institute for Science of Science (China). He received his doctorate majoring in history of science in 2011 from Shanghai Jiao Tong University, including two years of joint training in the University of Pennsylvania (2008-2010). His research focus is on new scientific and technological revolutions, and industrial transformation, especially the social impact and governance of AI. He regularly participates in the formulation of Chinese government strategies and plans on AI.

**Monique Mann, PhD,** is the Vice Chancellor's Research Fellow in Technology and Regulation at the Faculty of Law Queensland University of Technology (Australia). She is an Adjunct Researcher with the Law, Science, Technology and Society (LSTS) Research Centre at Vrije Universiteit Brussel (Belgium). Dr Mann is advancing a programme of socio-legal research on the intersecting topics of algorithmic justice, police technology, surveillance, and transnational online policing. She is author of *Politicising and Policing Organised Crime* (Routledge 2019), co-author of *Biometrics, Crime and Security* (Routledge 2018), and co-editor of *Good Data* (Institute of Network Cultures 2019). She is on the Board of Directors of the Australian Privacy Foundation where she Chairs the Surveillance Committee.

**Vidushi Marda** is a lawyer based in Bangalore (India). She leads ARTICLE 19's global research and engagement on artificial intelligence and human rights, and also works on strengthening human rights considerations in internet infrastructure. She is currently studying the impact of machine learning systems on the exercise of rights in Asia. Vidushi has produced case studies on the use of big data in governance in the global south, and has worked on understanding data flows within India's biometric identity project, *Aadhaar*. Her research has been cited by the Supreme Court of India in its seminal ruling on the right to privacy in 2017, the UK House of Lords Select Committee on AI and the UN Special Rapporteur on the right to freedom of opinion and expression. Vidushi is additionally a non-resident analyst on AI and rights with Carnegie India.

**Ben Wagner, PhD,** is an Assistant Professor and Director of the Privacy & Sustainable Computing Lab at Vienna University of Economics and Business (Austria). He previously worked at the Technical University of Berlin, European University Viadrina and the University of Pennsylvania. He is part of the Privacy & Usability training network and a member of the ENISA PSG advisory board. His research focuses on technology, human rights and accountable information systems. Ben holds a PhD in Political and Social Sciences from European University Institute in Florence (Italy).

**Saskia Witteborn, PhD,** is Associate Professor in the School of Journalism and Communication at the Chinese University of Hong Kong. She received her PhD from the University of Washington (US) and specializes in transnational migration and technologies. Saskia has worked with migrant groups in North America, Europe, and Asia and has strong expertise in forced migration, with contributions to the political economy of mobility and space, digital heterotopia, emotions, and grouping processes. Her work has appeared in journals such as the *Journal of Communication*, *Cultural Studies*, and the *Journal of Refugee Studies* as well as in edited collections. As Associate Director of the Research Centre on Migration and Mobility at CUHK, Saskia is generally interested in culturally grounded approaches to communicative action around agendas for change.

**Wei Wang** is a PhD candidate at the University of Hong Kong Faculty of Law. His research interests include IP & IT Law, Innovation & Competition Policy, particularly by means of computational legal studies. He engages as an Administrative Officer for the Creative Commons Hong Kong Chapter, of which he was also among the founding members. Prior to his PhD studies, Wei worked as Data Analyst of IP litigations in Shenzhen and Research Associate with a focus on AI and Copyright at Law and Technology Centre, HKU. Wei recently completed the WIPO/QUT LLM in Intellectual Property with a Dean's Scholarship. He also holds BEng, LLB and MPhil from Huazhong University of Science and Technology (China), with the support of the Research Center for the Judicial Protection of Intellectual Property approved by the Supreme People's Court of P.R.China. Wei is now a Member of the International Association of Artificial Intelligence and Law, Asian Privacy Scholars Network and Creative Commons Global Network.

# 1. Introduction

Artificial intelligence (AI) is a technology which is increasingly being utilised in society and the economy worldwide, and its implementation is planned to become more prevalent in coming years. AI is increasingly being embedded in our lives, supplementing our pervasive use of digital technologies. But this is being accompanied by disquiet over problematic and dangerous implementations of AI, or indeed, even AI itself deciding to do dangerous and problematic actions, especially in fields such as the military, medicine and criminal justice. These developments have led to concerns about whether and how AI systems adhere, and will adhere to ethical standards. These concerns have stimulated a global conversation on AI ethics, and have resulted in various actors from different countries and sectors issuing ethics and governance initiatives and guidelines for AI. Such developments form the basis for our research in this report, combining our international and interdisciplinary expertise to give an insight into what is happening in Australia, China, Europe, India and the US.

## What is AI?

Artificial Intelligence (AI) is an emerging area of computer science. There are numerous definitions and various terms used interchangeably to describe 'AI' within the academic literature (and also popular discourse) - these include, for example: algorithmic/profiling, automation, (supervised/unsupervised) machine learning, deep neural networks etc.

In general terms, AI could be defined as technology that automatically detects patterns in data, and makes predictions on the basis of them. It is a method of inferential analysis that identifies correlations within datasets that can, in the case of profiling, be used as an indicator to classify a subject as a representative of a category or group (Hildebrandt 2008; Schreurs et al 2008). A broad distinction is made between 'narrow' and 'general' or 'broad' AI. Narrow AI is an AI application which is designed to deal with one particular task and reflects most currently existing applications of AI in daily life, while general or broad AI reflects human intelligence in its versatility to handle different or general tasks. In this report when we discuss AI we refer to AI in its narrow form.

There are numerous applications of AI in a range of domains, perhaps contributing to definitional complexity, for example, predictive analytics (such as recidivism prediction in criminal justice contexts, predictive policing, forecasting risk in business and finance), automated identification via facial recognition etc. Indeed, AI has been deployed in a range of contexts and social domains, with mixed outcomes, including insurance, finance, education, employment, marketing, governance, security, and policing (see e.g., O'Neil 2016; Ferguson 2017).

## AI and Ethics

At this relatively early stage in AI's development and implementation, the issue has arisen of AI adhering to certain ethical principles (see e.g. Arkin 2009; Mason 2017), and the ability of existing laws to govern AI has emerged as key as to how future AI will be developed, deployed and implemented (see e.g. Leenes & Lucivero 2015; Calo 2015; Wachter et al. 2017a).

While originally confined to theoretical, technical and academic debates, the issue of governing AI has recently entered the mainstream with both governments and private companies from major geopolitical powers including the US, China, European Union and India formulating statements and policies regarding AI and ethics (see e.g. European Commission 2018; Pichai 2018).

A key issue here is precisely what are the ethical standards to which AI should adhere? Furthermore, the transnational nature of digitised technologies, the key role of private corporations in AI development and implementation and the globalised economy gives rise to questions about which jurisdictions/actors will decide on the legal and ethical standards to which AI may adhere, and whether we may end up with a 'might is right' approach where it is these large geopolitical players which set the agenda for AI regulation and ethics for the whole world.

Further questions arise around the enforceability of ethics statements regarding AI, both in terms of whether they reflect existing fundamental legal principles and are legally enforceable in specific jurisdictions, and also the extent to which the principles can be operationalised and integrated into AI systems and application in practice.

## What does 'ethics' mean in AI?

Ethics is seen as a reflection theory of morality or as the theory of the good life. A distinction can be made between fundamental ethics, which is concerned with abstract moral principles, and applied ethics (Höffe 2013). The latter also includes ethics of technology, which contains in turn AI ethics as a subcategory. Roughly speaking, AI ethics serves for the self-reflection of computer and engineering sciences, which are engaged in the research and development of AI or machine learning. In this context, dynamics such as individual technology development projects, or the development of new technologies as a whole, can be analyzed. Likewise, causal mechanisms and functions of certain technologies can be investigated using a more static analysis (Rahwan et al. 2019). Typical topics are self-driving cars, political manipulation by AI applications, autonomous weapon systems, facial recognition, algorithmic discrimination, conversational bots, social sorting by ranking algorithms, and many more (Hagendorff 2019).

Key demands of AI ethics relate to aspects such as the reflection of research goals and purposes, the direction of research funding, the linkage between science and politics, the security of AI systems, the responsibility links underlying the development and use of AI technologies, the inscription of values in technical artefacts, the orientation of the technology sector towards the common good, and much more (Future of Life Institute 2017).

Last but not least, AI ethics is also reflected within the framework of metaethics, in which questions about the effectiveness of normative demands are investigated. Ethical discourses can either be held with close proximity to their designated object, or it can be the opposite. The advantage of a close proximity is that those ethical discourses can have a concrete impact on the course of action in a particular organization dealing with AI. The downside is that this kind of ethical reflection has to be quite narrow and pragmatic. Uttering more radical demands only makes only sense when ethical discourses have a certain distance to their designated object. Nevertheless, those ethical discourses are typically rather inefficient and have hardly any effect in practice.

Another dimension of AI ethics concerns the degree of its normativity. Here, ethics can oscillate between irritation and orientation. Irritation equals weak normativity. This means an abstinence from strong normative claims. Instead, ethics just uncovers blind spots or describes hitherto underrepresented issues. Orientation, on the other hand, means strong normativity. The downside of making strong normative claims is that they provoke backfire- or boomerang-effects, meaning that people tend to react to perceived external constraints on action with that kind of behaviour they are supposed to refrain from.

Therefore, AI ethics must satisfy two traits in order to be effective. First, it should use weak normativity and should not universally determine what is right and what is wrong. Second, AI ethics should seek close proximity to its designated object. This implies that ethics is understood as an inter- or transdisciplinary field of study, that is directly linked to the adjacent computer sciences or industry organizations, and that is active within these fields.

## This Report

In this Report we combine our interdisciplinary and international expertise as researchers working on AI policy, ethics and governance to give an overview of some of our countries and regions' approaches to the topic of AI and ethics. We do not claim to present an exhaustive account of approaches to this issue internationally, but we do aim to give a snapshot of how some countries and regions, especially 'large' ones like China, Europe, India and the United States are, or are not, addressing the topic. We also include some initiatives at national level of EU Member States (Germany, Austria and the United Kingdom) and initiatives in Australia, all of which can be considered 'smaller'. The selection of these countries and regions has been driven by our own familiarity with them from prior experience.

We acknowledge the limitations of our approach, that we do not have contributions regarding this issue from Africa, Latin America, the Middle East, Russia, Indigenous views of AI and AI and ethics approaches informed by religious beliefs (see e.g. Cisse 2018; ELRC 2019; Indigenous AI n.d.). In future work we hope to be able to cover more countries and approaches to AI ethics.

We have specifically looked to government, corporate and some other initiatives which frame and situate themselves in the realm of 'AI governance' or 'AI ethics'. We acknowledge that other initiatives, such as those relevant to 'big data' and the 'Internet of Things' may also be relevant to AI governance and ethics; but with a few exceptions, these are beyond the scope of this report. Further work should be done on 'connecting the dots' between some predecessor digital technology governance initiatives and the current drive for AI ethics and governance.

The fast-moving nature of this topic and field is our reason for publishing this report in this current form. We hope the report is useful and illuminating for readers.

We welcome comments and other feedback on the work to date, and expressions of interest in future collaboration with us on this and related topics.

Please send any feedback to <u>angelacdalyATgmail.com</u>.

# 2. Global Level

## International organisations

At the international level, the most prominent AI ethics guidelines are the recently-released OECD Principles on AI (2019). The OECD is an international economic organisation of 36 Member States, mostly comprising high-income economies. The Principles have been endorsed by, among others, the US Trump Administration (Pressman & Lashinsky 2019), and six non-member states (Argentina, Brazil, Colombia, Costa Rica, Peru and Romania). In June 2019 ministers from the Group of 20 (G20) major economies agreed on a set of guiding principles for using AI, which are derived from the aforementioned OECD Principles, but are also characterised as non-binding (G20 2019). Notably, China and Russia are G20 countries but not OECD members (Koizumi 2019).

There are various activities that the United Nations (UN) and its constituent bodies is undertaking which relate to AI (ITU 2018). While at the time of writing the UN and its constituent bodies have not issued their own ethics or governance principles on AI, UNICEF and the United Nations Development Program (UNDP) are members of the multistakeholder Partnership on AI (discussed below). UNESCO is also working on a possible 'normative instrument' on the topic (ITU 2018). The United Nations Interregional Crime and Justice Research Institute is in the process of opening a Centre for Artificial Intelligence and Robotics in The Hague, Netherlands. An attempt in 2018 to open formal negotiations to reform the UN Convention on Certain Conventional Weapons to govern or prohibit fully autonomous lethal weapons were blocked by the US and Russia, among others (Delcker 2018).

The Council of Europe has also been active on the topic of AI. These developments are included in the next section on Europe.

At the 40th International Conference of Data Protection & Privacy Commissioners (ICDPPC) in 2018, which took place in Brussels, a Declaration on Ethics and Data Protection in Artificial Intelligence was released by delegates from various national data protection and privacy authorities. The Declaration sets out six guiding principles and calls for 'common governance principles on artificial intelligence' to be established. The ICDPPC has also set up a permanent working group on Ethics and Data Protection in Artificial Intelligence.

## Technical initiatives

The most prominent initiative from the technical community can be found in the IEEE's work on AI, in the form of its Global Initiative on Ethics of Autonomous and Intelligent Systems. The IEEE (Institute of Electrical and Electronic Engineers) has involved its membership of technical experts, but also reached beyond to non-IEEE members to participate in this project. The initiative has produced two versions to date of *Ethically Aligned Design*, involving 'hundreds of participants over six continents' (IEEE 2018). Version 2 includes five General Principles to guide the ethical design, development and implementation of autonomous and intelligent systems. In line with IEEE's general activities, the development of technical standards based on these discussions on ethics is envisaged by the Global Initiative, and a series of working groups have been set up under the Global Initiative to work towards this goal.

## Global multistakeholder initiatives

Some multinational corporations have also released their own ethics statements. Since many of these corporations originate in the US, they are included later in the section on the US. There is one group which may be considered truly global, and also multistakeholder in its membership, namely the Partnership on AI. As mentioned, some UN agencies are among its members, as well as NGOs (such as Article 19), academic research institutes (such as the Australian National University 3Ai Centre), public sector agencies (including the BBC) and also technology firms such as Amazon but also Chinese giant Baidu. The Partnership on AI has released its 8 'Tenets' (Partnership on AI n.d.).

The World Economic Forum, funded by its member corporations from around the world, has commenced various activities on AI, principally through its Center for the Fourth Industrial Revolution in San Francisco (US). Part of this Center's work is to co-design and pilot policy and governance frameworks including for AI with governments and corporations. In 2019 the WEF released a White Paper on the topic of AI governance.

# 3. Europe

In this section some AI governance and ethics initiatives which have been developed in Europe will be outlined. These include developments by the European Union (EU), Council of Europe (CoE) and in some individual nations (which happen to be members of both the EU and CoE).

## European Union

The EU has been positioning itself as a frontrunner in the global debate on AI governance and ethics. A major piece of legislation, the General Data Protection Regulation (GDPR) came into effect in 2018, and has a scope which extends to some organisations outside of the EU in certain circumstances. Of direct interest for AI governance are the provisions contained in Section 5 of the GDPR on the Right to Object (Article 21) and Automated Individual Decision-Making Including Profiling (Article 22). There is significant discussion as to precisely what these provisions entail in practice regarding algorithmic decision-making, automation and profiling and whether they are adequate to address the concerns that arise from such processes (see e.g. Edwards & Veale 2017; Wachter, Mittelstadt & Floridi 2017b).

Among other prominent developments in the EU is the European Parliament Resolution on Civil Law Rules on Robotics from February 2017. While the Resolution is not binding, it expresses the Parliament's opinion, and makes various requests of the European Commission to carry out further work on the topic. In particular, the Resolution 'consider[ed] that the existing Union legal framework should be updated and complemented, where appropriate, by guiding ethical principles in line with the complexity of robotics and its many social, medical and bioethical implications' and set out in its Annex a proposed Code of Ethical Conduct for Robotics Engineers, Code for Research Ethics Committees, Licence for Designers and Licence for Users. The Parliament also requested the European Commission to submit a 'proposal for a legislative instrument on legal questions related to the development and use of robotics and AI foreseeable in the next 10 to 15 years, combined with non-legislative instruments such as guidelines and codes of conduct as referred to in recommendations set out in the Annex'. At the time of writing, the Commission has not yet released such a proposal.

Further initiatives have occurred subsequent to this European Parliament Resolution. In March 2018, the European Commission issued a Communication on Artificial Intelligence for Europe, in which the Commission set out 'a European initiative on AI' with three main aims: of boosting the EU's technological and industrial capacity, and AI uptake; of preparing for socio-economic changes brought about by AI (with a focus on labour, social security and education); and of ensuring 'an appropriate ethical and legal framework, based on the Union's values and in line with the Charter of Fundamental Rights of the EU'.

Also in March 2018, the European Group on Ethics in Science and New Technologies, an independent advisory body to the President of the European Commission comprising interdisciplinary experts, released its Statement on Artificial Intelligence, Robotics and Autonomous Systems. The Statement proposed 'a set of basic principles and democratic prerequisites, based on the fundamental values laid down in the EU Treaties and in the EU Charter of Fundamental Rights'.

Most prominent of the EU initiatives has been the European Union High-Level Expert Group on Artificial Intelligence (a multi-stakeholder group of 52 experts from academia, civil society and industry) finalising its Ethics Guidelines for Trustworthy AI in April 2019 (2019a). They include 7 key, but non-exhaustive, requirements that AI systems should meet in order to be 'trustworthy'. The requirements will go through a 'piloting process' whereby they will be tested in private and public sector organisation, with feedback sought to inform a public document scheduled for release in early 2020.

A member of the High-Level Expert Group, Thomas Metzinger, criticised the process and output as 'ethics washing' in an op-ed for German newspaper Der Tagesspeigel in 2019. In particular he pointed to the removal of 'red line' 'non-negotiable' text from the final version of the Guidelines as an example of this, and called for academia and civil society to take charge of the discussion on AI governance and ethics, especially away from industry. Nevertheless, Metzinger still considers that the ethics guidelines produced by the Group are 'the best in the world' especially as compared to efforts from the US and China.

This 'first deliverable' of the High-Level Expert Group was followed by their 'second deliverable', Policy and Investment Recommendations for Trustworthy AI in June 2019 (2019b). The document contains 33 recommendations 'that can guide Trustworthy AI towards sustainability, growth and competitiveness, as well as inclusion – while empowering, benefiting and protecting human beings'. Among the recommendations, along with ones pertaining to education, research, government use of AI and investment priorities, is strong criticism of both state and corporate surveillance using AI, including that governments should commit not to engage in mass surveillance and the commercial surveillance of individuals including via 'free' services should be countered. This is furthered by a specific recommendation that AI-enabled 'mass scoring' of individuals be banned. The Panel also recommends that sustainability be taken account of, including the enactment of a circular economy plan for digital technologies and AI. The Panel calls for more work to be done to assess existing legal and regulatory frameworks to discern whether they are adequate to address the Panel's recommendations or whether reform is necessary in order to do so, with particular regard being paid to: the monitoring and restriction of automated lethal weapons; the monitoring of personalised AI systems built on children's profiles; and the monitoring of AI systems used in the private sector which significantly impact on human lives, with the possibility of introducing further obligations on such providers.

The language of 'red lines' is included in this document, and as mentioned above, the Panel expresses concern with some particular uses of AI, including examples it believes should be prohibited. This may stymie some of the previous criticism regarding the Guidelines being 'ethics washing' but it is still significant that that language was excluded from the Guidelines even if it has ended up in the Recommendations. Furthermore, it is unclear to what extent the Panel's Recommendations will actually be followed by EU institutions and put into practice in reality.

## Council of Europe

The Council of Europe (CoE), which includes all EU Member States as well as additional non-EU members in eastern Europe, Turkey and Russia, has also been active on the topic of AI.

Of these activities, there are two which directly relate to AI governance and ethics. The first is the European Commission for the Efficiency of Justice (CEPEJ) European Ethical Charter on the use of artificial intelligence (AI) in judicial systems and their environment, adopted in December 2018, which contains five principles to guide the development of AI tools in European judiciaries. The European Committee on Legal Co-operation (CDCJ) is at the time of writing working on draft guidelines for policymakers designing online dispute resolution systems (ODRs) to ensure compatibility with the right to a fair trial and the right to an effective remedy under the European Convention on Human Rights. These guidelines are projected to be released in late 2020.

The second notable CoE activity is the Guidelines on Artificial Intelligence and Data Protection published by the Consultative Committee of the Convention for the Protection of Individuals with regard to Automatic Processing of Personal Data (Convention 108) in January 2019. This follows Guidelines on Big Data issued in 2017, and the modernisation of Convention 108 which included additions to address algorithmic decision-making. Convention 108 includes among its signatories some non-CoE members including Mauritius, Mexico and Senegal.

## Germany

As a consequence of the significant financial support Germany is giving to AI research, a national 'AI Strategy' has been published (Bundesministerium für Bildung und Forschung et al 2018). The aims of the initiative are to strengthen Germany as a research location and to support the domestic economy. Just to give one example, between Tuebingen and Stuttgart the so-called 'Cyber Valley' is supposed to become one of the world's leading research locations for AI as a 'key technology'. The state government, companies, universities and other research institutions are cooperating in this project. The high monetary expenditure for AI research – the Federal Government will spend €500 million as a first step and €3 billion altogether – is justified almost exclusively with reference to the aim of survival in international competition or at least not wanting to fall behind, especially the US and China. According to the strategy, Germany is to become one of the 'world's leading locations for AI'.

Within the competitive relationships with other countries, Germany – in accordance with the principles of the EU Strategy for Artificial Intelligence (Pekka et al 2018) – intends to position itself in such a way that it sets itself apart from other, non-European nations through data protection-friendly, trustworthy, and 'human centered' AI systems, which are supposed to be used for the common good as well as for 'lighthouse applications' in the fields of climate and environment protection. At the centre of these claims is the establishment of the 'Artificial Intelligence Made in Germany' brand, which is supposed to become a globally acknowledged label of quality.

Part of this 'brand' is the idea that AI applications made in Germany, or, to be more precise, the datasets these AI applications use, stand under the umbrella of data sovereignty, informational self-determination and data safety. Moreover, to ensure that AI research and innovation is in line with ethical and legal standards a Data Ethics Commission was founded, which is able to recommendations to the Federal Government and to give advice on how to use AI in an ethically sound manner. The crucial question, however, is whether the tenets of AI ethics are implemented into practice effectively.

The federal government's strategy is to create one hundred new professorships in AI and Machine Learning. A total of twelve agglomerations for research and innovation are to be established. It is hoped that this will attract excellent researchers from abroad. Furthermore, it is intended that close cooperation with France will result in competitive advantages in research and industry. In addition, there is a special focus on the promotion of medium-sized enterprises. Further fields of action include the promotion of procedures to facilitate the auditing and interpretability of algorithmic prediction and decision-making systems as well as AI safety. Overall, it can be stated that Germany has a clear national 'roadmap' for the promotion and use of AI technologies.

## Austria

AI is viewed in Austria as offering a considerable competitive advantage to the nation. Drafting on an 'Artificial Intelligence Mission Austria 2030' was started, which has included a long list of stakeholder meetings to ensure participation of all relevant actors. While both participation methods used and stakeholder selection have not always been ideal, the initiative does represent at least an attempt to co-develop an Austrian AI strategy with hundreds of stakeholders. Its focus on the year 2030 is also evidently modelled on China's New Generation Artificial Intelligence Development Plan 2030 strategy, and like many similar European initiatives can be seen in response to China's position in the race to dominate the field of AI.

Austria has also shown a strong interest in European collaboration in this area, attempting to ensure that a frequently debated European 'Algorithms Rating Agency' or 'AI Ethics Authority' - modelled on the IAEA that is being currently being discussed in European policy circles - is eventually located in Vienna. Austria is the seat of numerous relevant international organisations such as the UN bodies, OSCE or the IAEA and sees such an authority as a natural continuation of its existing role in this area.

At the same time members of the Austrian government have openly expressed interest in creating a large national data pool, whereby Austrian citizens' data would be sold to the highest bidder in order to attract cutting edge data-driven research to Austria. Despite evident conflict with the GDPR and other existing data protection rules, this idea remains popular in relevant policy circles. It all stems from the acknowledgement that Austria knows that it is a small country and thus cannot compete at a global level in all domains. The data-pooling strategy is thus seen as a key competitive advantage to ensure that Austria is able to compete in a competitive international environment as a small country.

Due to the disintegration of the Austrian government as part of the 'Ibiza scandal' in May 2019 - involving a video featuring political bribes, a table of what looks like cocaine and attacks on political attacks on leading Austrian newspapers by the far right-wing FPÖ party - it is unclear how and even whether the Artificial Intelligence Mission Austria 2030 will continue. However, it is to be assumed that some version of this strategy will be implemented in the coming years, regardless of which government is in power.

## United Kingdom

The UK Government has linked AI development directly to its industrial strategy, and also seems to view this as giving the UK a potential competitive edge, particularly in the current context of the UK leaving the EU, or 'Brexit', and uncertainty as to what kind of political, economic and social future may lie for the UK subsequent to this point.

In its 2017 Industrial Strategy, the UK Government identified 'putting the UK at the forefront of the artificial intelligence and data revolution' as one of four 'Grand Challenges' for the country, adding that it would invest in business, research and education in the UK to meet the challenge, including an AI Sector Deal which was commenced in 2018. The UK government also proclaimed its vision that the UK 'will lead the world in safe and ethical use of data and artificial intelligence giving confidence and clarity to citizens and business', including by setting up a Centre for Data Ethics and Innovation as an advisory body to this effect in what the Government claims is a 'world first' (although this seems very similar to Germany's approach detailed above). The Centre has since been set up but has not produced any substantive outputs at the time of writing. As regards international governance, the Government asserted in its Industrial Strategy that it would be an 'active participant' in standard setting and regulatory bodies especially for AI and data protection.

The UK Parliament has also been active in its consideration of AI governance and ethics issues. An All-Party Parliamentary Group on AI was set up in 2017; and a Select Committee on AI was also formed to investigate the topic, seek input from interested stakeholders and then issued a report in 2018. The Select Committee's report (re)asserted the UK's place among the 'best countries in the world for researchers and businesses developing AI' but acknowledged that the UK may be unable to compete with the size of investments in AI being made by China and the US - although it could find better comparisons in the form of Canada and Germany. The UK could draw on what was perceived as its existing strengths and position itself as a leader in the ethical development of AI. The Select Committee did not consider that it was necessary to introduce AI-specific regulation at this point in time, but advocated for further work to be done assessing whether additions to existing legal and regulatory frameworks to deal with AI may be necessary in the future. The Select Committee did advance 5 non-legally binding 'overarching principles', as the basis for a possible cross-sector 'AI Code' that it suggested be formulated and developed by the Centre for Data Ethics and Innovation. Finally of relevance here is the Select Committee's recommendation to the UK Government to convene a 'global summit' by the end of 2019 involving different stakeholder groups to 'develop a common framework for the ethical development and deployment of artificial intelligence systems' which 'should be aligned with existing international governance structures'. It is unclear whether such an event will indeed be organised by the end of 2019.

In addition, the UK Government was the first government to partner with the World Economic Forum's aforementioned Center for the Fourth Industrial Revolution in its project to co-design guidelines for the public procurement of AI products and services for public sector uses (Russo 2018). At the time of writing, this work is underway, but no outputs have yet been produced.

Clouding the picture for the UK, in AI and other matters, is its pending departure from the European Union and the uncertainty about what the post-EU future holds for the country. The UK Government's activities in this area have been criticised by some stakeholder groups for not investing enough compared to other Western European countries and also making 'wrong' investments (Walker 2018). In addition, EU investment in AI and robotics will likely be no longer available for research and development in the UK, which to date has been a major recipient of these funds (Walker 2018). The EU also seems to be forging ahead of domestic UK initiatives regarding ethical AI development, which also may call into question how successful the desire to be a world leader, especially in ethical AI development, will be for the UK in a post-Brexit future.

# 4. India

India's approach to AI is substantially informed by three initiatives at the national level. The first is Digital India, which aims to make India a digitally empowered knowledge economy. The second is Make in India, under which the Government of India is prioritising AI technology designed and developed in India, and the third is the Smart Cities Mission (Marda 2018).

Alongside this, there is significant investment towards research, development and training in emerging technologies in particular from the Union Government. An AI Task Force constituted by the Ministry of Commerce and Industry in 2017 looked at AI as a socio-economic problem solver at scale. In its Report (Government of India Ministry of Commerce and Industry 2018) it identified 10 key sectors in which AI should be deployed, including national security, financial technology, manufacturing and agriculture, among others. Similarly, a National Strategy for Artificial Intelligence was published in 2018 (Niti Aayog 2018) that went further to look at AI as a lever for economic growth, social development, and considers India as a potential 'garage' for AI applications. While ethics are mentioned in both documents, they fail to meaningfully engage with issues of fundamental rights, fairness, inclusion, and the limits of data-driven decision making. These are also heavily influenced by the private sector, with civil society and academia, rarely, if ever, being invited into these discussions.

AI is being used in various sectors by private actors - from manufacturing, to healthcare, to finance. Notwithstanding encouraging developments, the current absence of data protection legislation in India raises crucial questions for how sensitive personal data is currently processed and shared. The current Personal Data Protection bill also fails to adequately engage with the question of inferred data, which is particularly important in the context of machine learning. India's biometric identity project, *Aadhaar,* could also potentially become a central point of AI applications in the future, with a few proposals for use of facial recognition in the last year, although that is not the case currently.

There is no ethical framework or principles published by the Government at the time of writing. It is likely that ethical principles will emerge shortly, following public attention on data protection law. Current references to AI are often in the context of data protection law, which is an increasing trend across jurisdictions.

# 5. China

Along with the EU, of the 'large jurisdictions' under consideration in this paper, China is the other one which has generated the most state-supported or -led AI governance and ethics initiatives.

In 2017 China's State Council issued The New-Generation AI Development Plan, which advanced China's objective of high investment in the AI sector in the coming years, and aim of becoming the world leader in AI innovation (FLIA 2017). An interim goal, by 2025, is to formulate new laws and regulations, and ethical norms and policies related to AI development in China. This includes participation in international standard setting, or even 'taking the lead' in such activities as well as 'deepen[ing] international cooperation in AI laws and regulations'.

Subsequent to this has been further initiatives on AI ethics and governance. In May 2019, the Beijing AI Principles were released by the Beijing Academy of Artificial Intelligence, which depicted the core of its AI development as 'the realization of beneficial AI for humankind and nature'. In addition, the Principles considered:

- the risk of human unemployment by encouraging more research on Human-AI coordination;

- avoiding the negative implications of 'malicious AI race' by promoting cooperation, also on a global level;

- integrating AI policy with its rapid development in a dynamic and responsive way by making special guidelines across sectors; and

- continuously making preventive and forecasting policy in a long-term perspective with respect to risks posed by Augmented Intelligence, Artificial General Intelligence (AGI) and Superintelligence.

The Principles have been supported by various elite Chinese universities and companies including Baidu, Alibaba and Tencent.

Another group comprising top Chinese universities and companies and led by the Ministry of Industry and Information Technology (MIIT)'s China Academy of Information and Communications Technology, the Artificial Intelligence Industry Alliance (AIIA), released its Joint Pledge on Self Discipline in the Artificial Intelligence Industry, also in May 2019 (Webster 2019). The Joint Pledge is, at the time of writing, open for comments from AIIA members and the general public until the end of June 2019 (Webster 2019). While the wording is fairly generic when compared to other ethics and governance statements Webster (2019) points to the language of 'secure/safe and controllable' and 'self-discipline' as 'mesh[ing] with broader trends in Chinese digital governance'.

Finally, an expert group formed of researchers at Chinese universities and established by the Chinese Government Ministry of Science and Technology released its eight Governance Principles for the New Generation Artificial Intelligence: Developing Responsible Artificial Intelligence in June 2019 (China Daily 2019). It has been reported that other experts, notably Kai-Fu Lee, made written submissions to the committee at earlier stages in their work (Laskai & Webster 2019). Again international cooperation is emphasised in the principles, including along with 'full respect' for AI development in other countries. A possibly novel inclusion is the idea of 'agile governance', that problems arising from AI can be addressed and resolved 'in a timely manner'. This principle reflects the rapidity of AI development and the difficulty in governing it through conventional procedures, for example through legislation which can take a long time to pass in China by which time the AI technology may have already changed. While 'agile policy-making' is a term also used by the EU High-Level Expert Panel, it is used in relation to e.g. the regulatory sandbox approach, as opposed to resolving problems, and is also not included in the Panel's Guidelines as a principle.

While, as mentioned above, Chinese tech corporations have been involved in AI ethics and governance initiatives both domestically in China and internationally in the form of the Partnership on AI, they also appear to be internally considering ethics in their AI activities. Tencent has its AI for Social Good programme and ARCC (Available, Reliance, Comprehensible, Controllable) Principles (Si 2018) but does not appear at the time of writing to have an internal ethics board to review AI developments.

However, the principles set by these initiatives so far lack legal enforcement/enforceability and policy implications - like the AI ethics/governance guidelines elsewhere.

## AI in Hong Kong

Under the 'One Country, Two Systems' framework, Hong Kong SAR remains a semi-autonomous region of China, with its own legal system until 2047. Its development, uptake and governance of AI present a different picture to that of mainland China.

In a regional report for the IEEE written in 2017, Yu pointed to AI adoption and development in Hong Kong being somewhat fragmented and only a little behind regional neighbours but viewed Hong Kong as having more progress to make 'before it can credibly tackle some of the legal, policy or ethical issues surrounding AI'.

Since that report was written, there has been one major development in Hong Kong. The Privacy Commissioner for Personal Data (PCPD), Hong Kong, issued an Ethical Accountability Framework in 2018, following industry consultation (but it is not clear if there was any consultation with academia and civil society). The discussion in the Framework explicitly links the issue of data ethics to AI, and acknowledges the additional guidance an ethical approach can give to the principles-based and technology-neutral legislation (the Privacy (Data Protection) Ordinance). The Framework includes a series of 'Enhanced Elements' and three 'recommended' Hong Kong Data Stewardship Values of 'Respectful, Beneficial and Fair' which were developed with the industry consultees, along with two assessment models for use by stakeholders. The Hong Kong Monetary Authority (2019) has encouraged 'authorised institutions' to adopt the Framework regarding personal data in the context of fintech development.

# 5. United States of America

Widely believed to rival only China in its domestic research and development of AI, the US has been less active institutionally regarding questions of ethics, governance and regulation compared to developments in China and the EU, until the recent Trump Administration Executive Order on Maintaining American Leadership in Artificial Intelligence from February 2019.

This Order has legal force, and creates an American AI Initiative guided by five high level principles and to be implemented by the National Science and Technology Council (NSTC) Select Committee on Artificial Intelligence. These principles include the US driving development of 'appropriate technical standards' and protecting 'civil liberties, privacy and American values' in AI applications 'to fully realize the potential for AI technologies for the American people'. Internationalisation is included with the view of opening foreign markets for US AI technology and protecting the US's critical AI technology 'from acquisition by strategic competitors and adversarial nations'.

Furthermore, executive departments and agencies that engage in AI related activities such as developing it, providing educational grants and 'regulat[ing] and provid[ing] guidance for applications of AI technologies' must adhere to six strategic objectives including protection of 'American technology, economic and national security, civil liberties, privacy, and values' and ensuring that technical standards for AI 'minimize vulnerability to attacks from malicious actors and reflect Federal priorities for innovation, public trust, and public confidence in systems that use AI technologies; and develop international standards to promote and protect those priorities'.

The Order also contains some Guidance on Regulation of AI Applications, whereby agencies are to receive a memorandum from the Office of Management and Budget within 180 days to inform their regulatory and non-regulatory approaches to AI which 'advance American innovation while upholding civil liberties, privacy, and American values' and shall 'consider ways to reduce barriers to the use of AI technologies in order to promote their innovative application while protecting civil liberties, privacy, American values, and United States economic and national security'. A draft of this memorandum is to be made publicly available before it is sent to these agencies, which appears not to have happened at the time of writing.

Finally, the National Institute of Standards and Technology (NIST) was tasked by the Executive Order with creating a plan for federal engagement in developing technical standards for reliable, robust, and trustworthy AI systems. In May 2019 NIST issued a Request For Information to this effect.

In addition, the US Department of Defense launched its AI Strategy, also in February 2019. The Strategy explicitly mentions US military rivals China and Russia investing in military AI 'including in applications that raise questions regarding international norms and human rights', as well as the perceived 'threat' of these developments to the US and 'the free and open international order'. As part of the Strategy, the Department asserts that it 'will articulate its vision and guiding principles for using AI in a lawful and ethical manner to promote our values', and will 'continue to share our aims, ethical guidelines, and safety procedures to encourage responsible AI development and use by other nations'.

The Department's Joint Artificial Intelligence Center will, among other tasks, '[f]acilitate AI planning, policy, governance, ethics, safety, cybersecurity, and multilateral coordination'. As regards more detailed principles, the Department asserted that it would develop principles for AI ethics and safety in defence matters after multistakeholder consultations, with the promotion of the Department's views to a more global audience, with the seemingly intended consequence that its vision will inform a global set of military AI ethics.

As well as these recent interventions from the federal government, the US has a stronger record of AI ethics and governance activity from the private and not-for-profit sectors. Various US-headquartered/-originating multinational tech corporations have issued ethics statements on their AI activities, notably the Microsoft AI Principles. DeepMind (part of Google's Alphabet group of companies) also has its own Ethics and Society Principles. In 2019, Google itself announced that it had set up an AI Ethics committee of experts to inform its AI activities, but not long after this announcement, Google disbanded the committee, seemingly due to the controversial views of one panel member and the negative public reaction and reaction from Google employees (Turner 2019).

There are various not-for-profit organisations and foundations based in the US which have also been active in AI governance and ethics discussions. The Future of Life Institute released its 23 Asilomar AI Principles, developed in conjunction with their Beneficial AI conference in 2017. Participants in the process appeared to come mainly from academia and industry. OpenAI, a mixed organisation with for-profit and not-for-profit wings, released its OpenAI Charter, detailing the principles the organisation uses to guide its activities in accordance with its mission that artificial general intelligence (systems which outperform humans at economically valuable work) benefits all humanity.

As regards applications of AI in the form of facial recognition technology, in May 2019 the city of San Francisco prohibited the use of facial recognition technology by city agencies and the police department - although the technology was seemingly not being used by the police department there prior to the ban (Sandler 2019). The prohibition was motivated by concerns about the inaccuracies in using facial recognition technology especially for people of colour, and about racial discimination in how facial recognition technology has been deployed elsewhere to target specific, particularly racial, groups and communities (Sandler 2019).

# 6. Australia

Australia is in a unique situation as the only Western democracy without comprehensive enforceable protection of human rights (that is, no bill of rights, no comprehensive constitutional protection of rights). Despite this, there has been increasing attention in Australia on the human rights impacts of technology, and the development of an ethics framework for AI. Specifically, the Australian Human Rights Commission has commenced a Technology and Human Rights project, including releasing a white paper for public consultation, although a final report is yet to be published. Further, the Australian Government Department of Industry, Innovation and Science has recently released a discussion paper to inform the development of Australia's ethics framework for Artificial Intelligence (initial submissions on the discussion paper closed at the end of May 2019). In addition, in 2018, the Office of the Victorian Information Commissioner released an issues paper on Artificial Intelligence and Privacy.

The most prominent of these developments is the proposed Australian Ethical Framework currently under development by Data 61 and CSIRO in the Commonwealth Department of Industry, Innovation and Science. The discussion paper commences with an examination of existing ethical frameworks, principles and guidelines. The report includes a selection of case studies, these are largely international or US based, which overshadows the unique Australian (e.g. socio-political) context.

The report dedicates a chapter to considerations of 'data governance' (i.e. privacy and data protection) which has been critiqued extensively by a coalition of Australian privacy experts, as representing a fundamental misunderstanding of Australian privacy law (Salinger 2019). Further, there is a focus on matters of 'consent' where it may not be relevant in either current or future data processing landscapes. In addition, when discussing issues of data governance there is a need to distinguish between personal information and sensitive information, and also to consider sensitive inferences (see Wachter and Mittelstadt's work (2018) on a right to reasonable inferences). The report sets forth a very narrow understanding of the negative impacts of AI for privacy, for example, a focus on data breaches or the potential for re-identification of de-identified data. Rather, there is a need to consider other harms, for example, such as those that could arise from automated decision-making. The report does indeed dedicate a chapter on automated decisions but does not consider or refer to regulatory approaches to respond or regulate to automated decision making, processing or profiling (for example, Article 22 of the EU's GDPR). Further, there is a need to consider the complexities of processing/profiling large sets of data, abstraction from data, including sensitive inferences, as mentioned above. The report then considers examples of AI in practice, and then outlines a proposed ethical framework.

## Principles for Australia's AI Ethical Framework

The development of the proposed Australian AI ethical framework was guided by a steering committee comprising industry, government, community organisations, and CSIRO/Data61 researchers. It should also be noted that CSIRO/Data61 is a Commonwealth entity that is developing AI technology for the Commonwealth government. The discussion paper sets out eight core or key principles to form an ethical framework for AI, namely: generates net-benefits; do no harm; regulatory and legal compliance; privacy protection; fairness; transparency and explainability; contestability; and explainability.

The proposed Australian AI ethical framework is accompanied by a 'toolkit' of strategies, which appear to be attempts to operationalise the high level ethical principles in practice. The 'toolkit' indicates how the high-level ethical principles are intended to translate into practice. These include: impact assessments; internal/external review; risk assessments; best practice guidelines; industry standards; collaboration; mechanisms for monitoring and improvement; recourse mechanisms, and; consultation.

Australia's ethical framework and the associated 'toolkit' is presently under development, and will continue to evolve and be refined on the basis of public submissions and consultation over the near future. In general, the proposed Australian principles do not seem very different to other ethical principles especially from other Western jurisdictions.

# 7. Reflections, issues and next steps

Here we offer some tentative reflections on the country/region profiles outlined above, and also the issue of the global governance of AI more generally which we set out here. Some of these reflections lead to further questions for further analysis of the different ethics and governance initiatives (and lack thereof in some cases).

## 1. Have and have-nots

Of the large jurisdictions we have considered, it is interesting to note that the European Union and China seem most advanced in their consideration of ethical aspects of AI. The US may soon make up for lost time subsequent to the directions in the Executive Order from earlier this year. However, India stands alone as the one of the four 'large jurisdictions' under consideration which has not yet developed ethics guidelines for AI.

## 2. Competition vs collaboration

Themes of competition loom large over national/regional AI policies, as regards competition with other 'large' countries or jurisdictions. It is widely believed and asserted that the US and China are the global forerunners in AI research and development and are in direct competition with each other (Cave and ÓhÉigeartaigh 2018). This may be reflected in the US Executive Order being framed around preserving the US's competitive position, and also the Chinese ambition for China to become the global AI leader in 2030. This competition takes both economic and also security (including military) dimensions, especially in the latter case for state-used AI and AI technologies applied in public infrastructure.

However, there are also calls for global collaboration on AI ethics and governance, notably from the Chinese initiatives. This can be contrasted, on the face of it at least, with language from the US government initiatives which seems to reflect the 'America First' approach of the Trump Administration more broadly, whereby ethics and governance may be devised in the US for AI and then 'shared' or exported to the rest of the world. While less explicit, the European approaches which reference European legal texts such as the Charter of Fundamental Rights of the EU may foresee an attempt to export 'European values' regarding ethical AI globally, as de facto may be happening for data protection via the EU GDPR's extraterritorial reach.

Not only among the big powers, but also internally do we also see some economic competition, particularly in the case of the EU where there are EU-level AI initiatives and strategies, but also possibly competing initiatives at the national level in some Member States. The relationship between Member States and the EU centrally may influence the development of such national strategies, especially when that relationship has broken down, as in the case of the UK leaving the EU. The UK's own national AI strategies and attempt to position the UK as a leader in ethical AI ought to be considered against this backdrop, especially when its European competitor Germany is also positioning itself in this space.

For smaller countries which may not be able to compete with the 'big boys' of global AI, Austria suggests that they may be willing to engage in less ethical projects to attract attention and investment (although this may also be the case for India if it positions itself as an AI 'garage'), and the Australian example shows how they may be 'followers' rather than 'leaders' inasmuch as they receive ethical principles and approaches formulated by other, similar but larger countries.

## 3. Similarities and differences

In many of the AI ethics/governance statements, we see similar if not the same concept reappear, such as transparency explainability, accountability etc. Hagendorff (2019) has pointed out that these principles, frequently encountered, are often 'the most easily operationalized mathematically' which may account partly for their presence in many initiatives.

Some form of 'privacy' or 'data protection' also features frequently, even in the absence of robust privacy/data protection laws as in the US example. In the case of India, a lack of data protection law at the time of writing is viewed as a reason explaining why there is no AI ethics/governance document yet issued there.

As mentioned earlier, some AI ethics initiatives present some differences, both in terms of which actors have formulated them (public vs private sector; civil society involvement; academic expertise etc), the content of principles articulated (e.g. China's 'agile governance') and the general framing of the statements of principles.

Nevertheless, behind some of these shared principles may lie different cultural, legal and philosophical understandings. This is a point to interrogate further in future work.

## 4. What's not included?

Another important issue is what is not included in AI ethics/governance initiatives, i.e. what is missing from the lists of principles. Is reference made to other government or corporate initiatives which may contradict the principles? To what extent are the 'hidden costs' of AI made visible and internalised, such as the energy and other human resources and raw materials needed for the systems (Hagendorff 2019)? The language of sustainability in the EU High-Level Expert Panel's Recommendations does acknowledge the environmental and sustainability aspects and concerns regarding the creation and use of AI technologies.

A further question is the extent to which the very use of AI in the first place is ethically interrogated by AI and ethics and governance statements, or just assumed to be used, or will inevitably be used.

And what about the broader social contexts in which AI finds itself deployed - this seems rarely if ever to be considered in AI ethics/governance initiatives explicitly.

## 5. What's already there?

There are already different areas of existing law, policy and governance which will apply to AI and its implementations including technology and industrial policy, data protection, intellectual property, fundamental rights, private law, administrative law, etc. Sometimes these existing regimes are taken account of in AI ethics/governance initiatives but in many cases, they are not, or indeed in the Australian example, may have been mischaracterised and misunderstood.

While less 'exciting' and 'novel' than proposing new governance frameworks, important work needs to be done to understand better the interactions of these existing frameworks with AI and the extent to which further action is necessary, or could be guided, as has been called for in the EU. It is important for those to whom AI ethics and governance guidelines area addressed to be aware that they may need to consider, and comply with, further principles and norms in their AI research, development and application beyond those articulated in AI-specific guidelines.

## 6. Implementation and enforceability

Almost all of the AI ethics and governance documents we have considered do not have the force of binding law. The US Executive Order is an exception in that regard, although constitutes more a series of directions to government agencies rather than a detailed set of legally binding ethical principles. This situation overall leads to concerns about 'ethics washing' (Wagner 2018; Watts 2019) as mentioned earlier in relation to the EU High-Level Expert Panel activities, but on a broader scale: that ethics and governance initiatives without the binding force of law are mere 'window dressing' while unethical uses of AI by governments and corporations continue. Indeed, in various of the countries and regions we have examined, despite the existence of ethical guidelines on AI, unethical AI applications exist, which may fail a test against those very principles.

This also leads to a broader question about the operationalisation of ethical and governance initiatives, especially those which articulate principles and norms for AI. Can they and will they actually be implemented into law, business and technical practices in public and private sector AI in their respective countries, regions and corporations? How will this be assessed? Will there be meaningful consequences if this does not happen?

Some ethics statements also include regulatory or legal compliance as an ethical principle, such as in the Australian example. While this may acknowledge that existing laws and regulations may not have been complied with in the past by AI researchers, developers and implementers, such an ethical principle may seem odd, inasmuch as it may be considered the law ought to be complied with anyway. However, such an ethical principle also does not account for situations where the existing law may be unethical or otherwise lacking.

A further issue arises around organisations or researchers from a particular country or region which does have AI ethics/governance principles 'jurisdiction shopping' to a location which does not or has laxer standards to research and develop AI with less 'constraints'? This offshoring of AI development to 'less ethical' countries may already be happening and is something that should in particular be addressed in national/regional ethics and governance initiatives.

A historical perspective is also warranted regarding the likelihood of success for AI ethics/governance initiatives, in the form of examining the success or otherwise of previous attempts to govern new technologies, such as biotech and the Internet, or to insert ethics in other domains such as medicine (see Mittelstadt 2019). While there are specificities for each new technology, different predecessor technologies from which it has sprung, as well as different social, economic and political conditions, looking to the historical trajectory of new technologies and their governance may teach us some lessons for AI governance and ethics.

## 7. Who is involved?

A key question for analysing AI ethics and governance activities further in future work is examining which actors are involved in their processes of formulation - at the global or national levels. To what extent are academic experts and civil society groups involved and to what extent are their voices and input heard and considered? Do civil society groups which are involved represent the public in practice, or subsections of the public? Who funds the groups and initiatives? Is there enough genuine public consultation to ensure AI has a social licence to operate in a particular country, or for particular applications? Is the formulation of AI ethics and governance largely a technocratic exercise? To what extent are participants in AI and its governance and ethics debates and initiatives demographically representative of the population at large? Hagendorff (2019) has criticised a lack of gender diversity in the AI field as an example of how ethical goals are being 'underachieved'.

We have identified scope for further work on AI ethics and global governance in the preceding paragraphs as well as earlier in this report when we outlined the limitations of this work. Further work may also be conducted to continue to track and analyse emerging and new AI ethics and governance initiatives, as well as appraise how existing initiatives are being implemented. We as a collective and as individual researchers may do some of this work ourselves, but we also look to such further work being done by others. To that extent, if you are interested in collaborating with some or all of us, please get in touch!

# Further Resources